# MyoGestic: EMG Interfacing Framework for Decoding Multiple Spared Degrees of Freedom of the Hand in Individuals with Neural Lesions

**Authors:** Raul C. Sîmpetru[1]†, Dominik I. Braun[1]†, Arndt U. Simon[1], Michael März[1], Vlad Cnejevici[1], Daniela Souza de Oliveira[1], Nico Weber[2], Jonas Walter[2], Jörg Franke[2], Daniel Höglinger[3], Cosima Prahm[3], Matthias Ponfick[4], and Alessandro Del Vecchio[1]*

**Affiliations:**

[1]Neuromuscular Physiology and Neural Interfacing Laboratory, Friedrich-Alexander-Universität Erlangen-Nürnberg; 91052 Erlangen, Germany

[2]Institute for Factory Automation and Production Systems, Friedrich-Alexander-Universität Erlangen-Nürnberg; 91054 Erlangen, Germany

[3]Department of Plastic and Reconstructive Surgery, BG Trauma Clinic, University of Tübingen; 72076 Tübingen, Germany

[4]Querschnittzentrum Rummelsberg, Krankenhaus Rummelsberg GmbH; 90592 Schwarzenbruck, Germany

*Corresponding author. Email: alessandro.del.vecchio@fau.de

†Co-first authors

**One Sentence Summary:** We developed a neural interface capable of decoding multiple spared degrees of freedom in individuals with motor impairments.

**Abstract:** Restoring limb motor function in individuals with spinal cord injury (SCI), stroke, or amputation remains a critical challenge, one which affects millions worldwide. Recent studies show through surface electromyography (EMG) that spared motor neurons can still be voluntarily controlled, even without visible limb movement. These signals can be decoded and used for motor intent estimation; however, current wearable solutions lack the necessary hardware and software for intuitive interfacing of the spared degrees of freedom after neural injuries. To address these limitations, we developed a wireless, high-density EMG bracelet, coupled with a novel software framework, MyoGestic. Our system allows rapid and tailored adaptability of machine learning models to the needs of the users, facilitating real-time decoding of multiple spared distinctive degrees of freedom. In our study, we successfully decoded the motor intent from two participants with SCI, two with spinal stroke, and three amputees in real-time, achieving several controllable degrees of freedom within minutes after wearing the EMG bracelet. We provide a proof-of-concept that these decoded signals can be used to control a digitally rendered hand, a wearable orthosis, a prosthesis, or a 2D cursor. Our framework promotes a participant-centered approach, allowing immediate feedback integration, thus enhancing the iterative development of myocontrol algorithms. The proposed open-source software framework, MyoGestic, allows researchers and patients to focus on the augmentation and training of the spared degrees of freedom after neural lesions, thus potentially bridging the gap between research and clinical application and advancing the development of intuitive EMG interfaces for diverse neural lesions.

**Main Text:**

**INTRODUCTION**

Living without a limb's motor function is the tragic reality for 20.64 million people worldwide with spinal cord injuries (SCI) (*1*), hundreds of thousands with spinal strokes (*2*, *3*), and 2.23 million with unilateral limb amputations (*4*).

Recovering limb function after these injuries remains a significant challenge. Numerous studies indicate that, despite the lack of visible movements, spared alpha motor neurons can still be voluntarily controlled in patients with motor complete SCI or amputation (*5–7*). Recently, we have used surface electromyography (sEMG), a non-invasive neural interface (*8*), to investigate individuals living with SCI who have been clinically labeled as having a motor-complete injury (*9*), and found that these patients can precisely control the activity of several spared motor units (MU) during intentional attempts to move the hand digits (*5*, *6*).

To date, no EMG-based decoding system can be used to recover limb motor function for multiple types of neural lesions (*10*). Current systems are high-end EMG equipment (e.g. Quattrocento, OT Bioelettronica S.r.l., Torino, Italy; RHD Recording System, Intan Technologies, Los Angeles, USA) that are made for laboratory settings, resulting in recording capabilities of multiple hundred channels (384 - Quattrocento; 1024 - RHD Recording System), however, at the expense of being big and impractical (130 x 395 x 271 mm - Quattrocento; 170 x 420 x 600 mm - RHD Recording System) for any daily usage. To counteract this, portable wireless systems have been developed that can record EMG signals in bipolar (*11*) or monopolar (*12–14*) derivation. However, these systems lack the modular software that is needed to interface the spared spinal cord output in an intuitive and effective manner such that myocontrol research may be iterated faster. As a result, current systems are used with custom-made software that can only be operated by experienced researchers and medical professionals (*5*, *6*, *15*, *16*). Most importantly, current software can only decode a limited number of degrees of freedom. The models are either not tailored specifically for individuals with impairments (*11*), or they are designed for specific participants but without any physiological a-priori constraints (*17*, *18*) making them less scalable for decoding multiple degrees of freedom intuitively for the impaired user.

To bridge the gap between research and clinical practice and create a generalizable high-density EMG interface, we developed a lightweight (76 ± 2 g, see Materials and Methods), wireless 32-channel monopolar EMG bracelet with a novel software framework, MyoGestic, which facilitates rapid adaptability to the needs of impaired participants (Fig. 1). Our software, capable of both high-performance classification and regression algorithms, allows immediate validation of new hand movements while keeping participants engaged and informed about their progress. The software is integrated with the input data from two digital hands, a hand that is controlled by the user (thereafter the *predicted hand*), and a *control hand*, that can be used by the experimenter to gather real-time data labels for the machine learning models.

Using the proposed non-invasive neural interface, we were able to decode for the first time multiple distinctive gestures in real-time from four spinal cord injured participants (n = 4 gestures, 2 spinal strokes and 2 traumatic lesions, Table 1) and three amputees (n = 5 gestures, 2 transradial and 1 transcarpal, Table 1, Video S1). Remarkably, we were able to achieve several controllable degrees of freedom (distinct EMG activity patterns that can be used as control signals) for all participants in less than 10 minutes after explaining the experiment (Video S2). The decoded degrees of freedom could then be interfaced either on a display as a controllable virtual hand, as input signals for a wearable orthosis/prosthesis, or as a 2D cursor (Fig. 1).

This framework could enable researchers to iterate on myocontrol algorithms more quickly, with a participant-centered approach that allows feedback to be integrated immediately and during a-posteriori (offline) analysis. By providing already tested code (e.g., real-time plotting or reactive user interface), researchers can dedicate more time to their algorithms and not be concerned with redesigning the experimental setup every time.

The software framework is available at the following link as open-source software: https://github.com/NsquaredLab/MyoGestic.

## RESULTS

### Design of a universal user-adaptable non-invasive neural interface system

We developed a non-invasive neural interfacing system that consists of a commercially available 32-channel monopolar EMG bracelet (Fig. 1, codeveloped with OT Bioelettronica S.r.l., Turin, Italy) and an open-source software framework designed for fast experimentation and adaptability to participant-specific feedback. The software is split into two interfaces, one for the participants (Fig. 1, large monitor) and one for the experimenter to trigger specific events (Fig. 1, laptop screen) such as the recording of a new attempted movement.

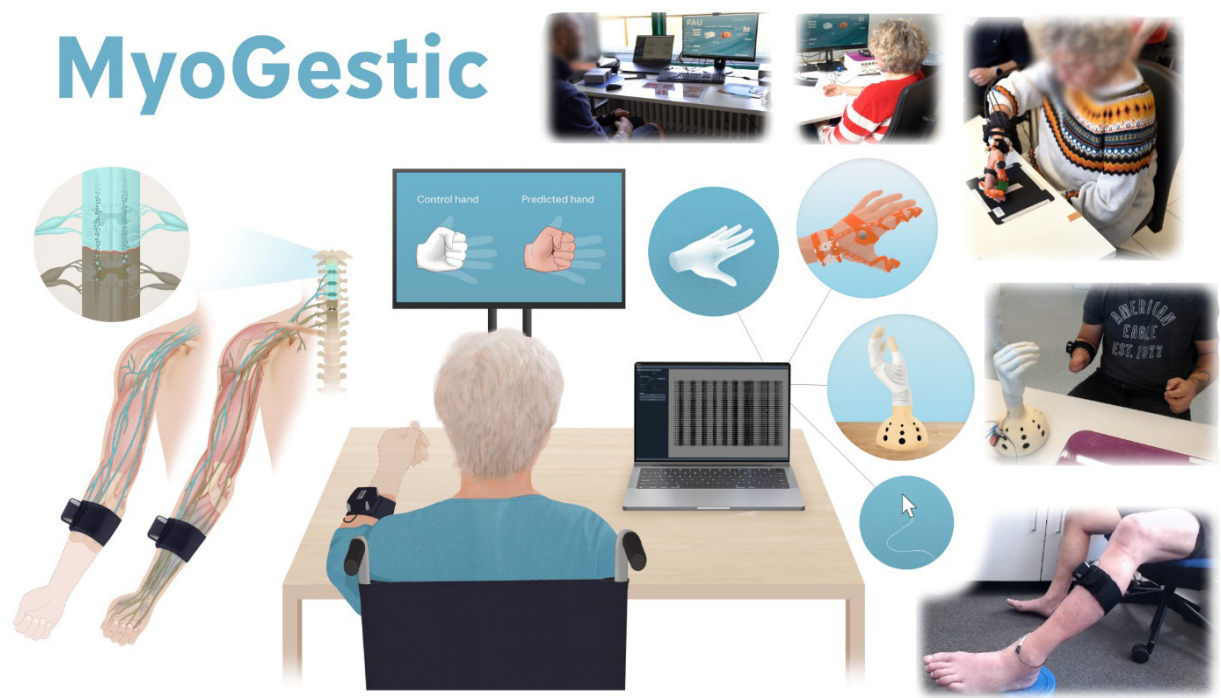

*Fig. 1. **Study overview.** Four individuals with spinal cord injuries (two with traumatic lesions and two with spinal strokes) and three amputees (two transradial and one transcarpal) attempted different movements displayed on a screen while their electrophysiological activity was recorded using a 32-channel wireless EMG bracelet. We developed a highly intuitive decoding system that enables participants to control a virtual hand, an orthosis, a prosthesis, or a cursor using their spared neural activity. The framework is highly adaptable to individual needs and was successfully used to decode four degrees of freedom in real-time, even for a participant who suffered a motor-complete (AIS B) spinal cord injury 21 years ago. The total time from explaining the experiment (to a naïve participant) to controlling four degrees of freedom of the hand in real-time can take less than 10 minutes. In a trained participant it takes less than 4 minutes to wear the EMG bracelet, train the machine learning model, and control the spared degrees of freedom.*

Individuals who suffered from chronic hand function loss (*5*, *6*, *15*) often struggle to imagine and attempt hand movements, which imposes a challenge in recalling or learning new neural commands necessary to execute them. To address this challenge, we designed the participant interface featuring two hands (Fig. 1, large monitor): one white hand guiding participants (the control hand) and one light skin-toned displaying their predicted movement (the predicted hand). This setup aided the participants in conceptualizing and maintaining a consistent frequency of specific hand movements that can be defined by the experimenter. By decoding directly the EMG activity associated with the control hand, we allowed the participants to intuitively attempt the natural motor commands that once controlled the movement of the hand (*6*). We pre-defined nine hand movements for our experiments (Fig. 2A): resting state with all digits extended; individual finger flexions; grasp; two-finger pinch (thumb and index); three-finger pinch (thumb, index, and middle). The state of each hand at any given time is represented by a 9D vector, whose elements control a simplified joint model of the hand (Fig. 2B, see Materials and Methods). This design choice enables straightforward definition of novel hand movements as well as rapid and accurate tracking of ongoing ones.

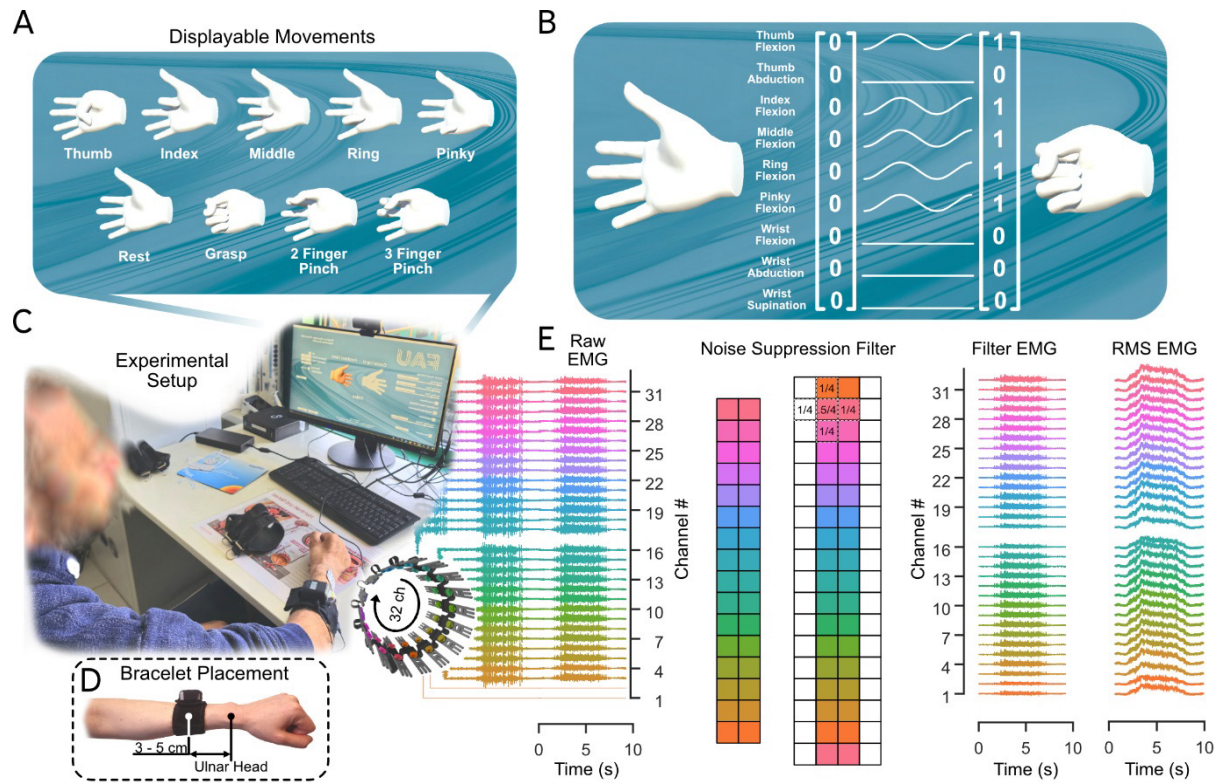

*Fig. 2. **Experimental setup and EMG processing.** (**A**) To assist participants in visualizing and maintaining a constant movement speed, we have pre-set nine distinct movements for display. These movements include the resting state, where the hand is in a neutral position with minimal activity, all five individual finger flexions, a hand grasp, and two types of pinches: between the thumb and index finger, and between the thumb, index, and middle finger. (**B**) The state of the virtual hand at any given time step is represented by a 9D vector with values ranging between 0 and 1. A value of 0 indicates that a particular joint is in its rest state, while a value of 1 indicates that the joint is fully activated. (**C**) The experimental setup was intentionally kept minimal, comprising a laptop running MyoGestic, an EMG bracelet, and a monitor. This design allowed the experiment to be conducted conveniently in different locations, such as hospitals (as exemplified here) or in our laboratory environment. The monitor showed two hands. The white "control" hand controls the participant's pace by demonstrating any of the pre-set movements discussed in panel A. Simultaneously, the predictions generated by our model are shown on the 'predicted' hand at 32 Hz. This frequency ensures smooth transitions between movements, mitigating the uncanny valley effect. (**D**) The 32-channel EMG bracelet was positioned for patient convenience at 3 to 5 cm from the distal ulnar head. The ground reference was placed on the nearest accessible bony structure, such as the ulnar head or elbow. (**E**) EMG processing pipeline example using data from a bracelet with 2 electrodes that had no skin contact (1 and 17). The initial EMG matrix is reshaped to represent the bracelet as the physical 16-row 2-column structure. Row-wise circular padding simulates the looping of the bracelet, followed by column-wise zero padding for applying a 3x3 high-frequency noise suppression kernel using convolution. Finally, one root mean squared (RMS) value is computed per channel.*

The lightweight (76 ± 2 g) EMG bracelet was fabricated from a flexible printed circuit board, featuring gold-coated copper electrodes housed within a fabric sleeve (Fig. 1, Fig. 2C; see Materials and Methods). Five bracelet lengths (18, 19, 21, 23, and 33 cm) were available, with the one closest to the participant's forearm circumference selected to ensure optimal EMG signal quality and minimal electrical noise interference. The selected bracelet was placed about 3 to 5 cm away from the wrist (distal ulnar head, Fig. 2D). An adhesive ground reference electrode was attached to the elbow. The signals were recorded in real-time at 2000 Hz and streamed as 111 non-overlapping EMG windows per second using a commercially available wireless amplifier (see Materials and Methods).

To achieve the highest possible prediction rate (111 Hz) and demonstrate the potential of wearable neural interfaces for individuals with hand motor impairments, we limited our signal processing to only what was necessary. The EMG signals were filtered to remove high-frequency noise such as movement artifacts, followed by computation of the root mean squared (RMS) value for each channel within each window (Fig. 2E; see Materials and Methods). We selected the CatBoost algorithm (*19*), an established model within the machine learning community, for the classifier, as it has been shown to excel among its peers in terms of performance, and benefits from being open sourced under an Apache 2.0 license. Moreover, the user can also choose to predict simultaneously and proportionally with a custom-made convolutional neural network from the raw monopolar EMG signals (*20, 21*).

Our interface, used to orchestrate the recording, training, and validation of decoded degrees of freedom, strikes a balance between streamlining processes and providing programmable freedom. Each step - connecting to the device, recording, and training/validation - is separated into panels that can be viewed one at a time, minimizing cognitive load during the experiment and reducing potential errors. Regardless of the current step, we display all the EMG channels in real-time, allowing us to monitor signal quality and intervene if necessary (e.g., addressing issues with bracelet slipping of the arm stump). The design minimizes the time it takes for participants to receive feedback, keeping them informed throughout each stage. Video S2 shows in real-time the workflow needed to record and control three degrees of freedom on a transcarpal amputee.

**Motor intent decoding from individuals with SCI**

Using our neural interface system, we recorded data from three participants with cervical (C5-C6) SCI (Table 1; see Materials and Methods). Participants 1 and 2 experienced a spinal ischemia (stroke), while participant 3 had a traumatic lesion. The participants were seated in front of a monitor and received a 10-minute introduction of the virtual hand interface (Fig. 2A-C) while the proposed EMG bracelet was fitted (Fig. 1; Fig. 2C & D; Fig. S1A). We visually inspected the signals for noise caused by issues such as poor electrode-skin contact due to the bracelet being too loose.

The next 15 minutes were dedicated to identifying the hand movements that the participants could reliably recall after years (12, 2, and 21 respectively) of living with hand paralysis (Fig. 3A). During each movement attempt, we visually screened the EMG signals for voluntary changes. After identifying three different movements in addition to the rest state that were controllable, we visualized, recorded, and trained the machine learning model with their data using MyoGestic (Fig. 2A-C; Fig. S1A) for 20 seconds per degrees of freedom. Participant 3 requested that their data recording duration to be adjusted to 30 seconds per movement, for personal preference. We were able for the first time to identify in real-time four distinguishable degrees of freedom by the participants with SCI.

Using this data, we trained a CatBoost (*19*) classifier model for each participant to decode the four degrees of freedom. All three participants were able to produce voluntary neural signals for grasping, flexing the thumb, and flexing the index finger individually.

To validate the trained classifier, we asked each participant to control the previously executed movements in real-time. Participants followed the guiding hand (Fig. 1; Fig. 2A), with model predictions displayed as a light skin-toned hand (Fig. 1; Fig. 3A). Each participant was asked to follow each degree of freedom for 6 repetitions within 45 seconds, although participant 3 opted to do more repetitions for personal preference.

Fig. 3B shows the predictions using both an uncertainty-unaware (naïve) approach and an uncertainty-aware approach (conformal prediction; see Materials and Methods), along with the synchronous EMG signals for the first participant. Briefly, we used conformal prediction to gauge uncertainty, thereby making the predictions of the machine learning model more stable by disregarding uncertain ones. For example, this safety feature helps the prevention of non-voluntary openings of the hand while attempting to grasp, but it may also reduce the user's reaction speed since the model must be certain that opening the hand is the intended action. Results for the other two participants are provided in Fig. S1A. Video S3 shows the third participant during their attempt at controlling four degrees of freedom in real-time. The participants achieved accuracies of $79.4 \pm 8.4\%$, $81.8 \pm 15.6\%$, and $75.8 \pm 1.5\%$ using the naïve approach, and $88.3 \pm 4.7\%$, $84.9 \pm 12.4\%$, and $56.4 \pm 25.7\%$ using conformal prediction (Fig. 3C). Notably, the third participant (21 years living with a motor-complete SCI, Table 1) had lower accuracy with the conformal prediction than with the naïve approach.

The time taken to record movements and validate control over the trained degrees of freedom was 4 min 12 s, 3 min 55 s, and 7 min 39 s for each participant, respectively (Fig. 3D).

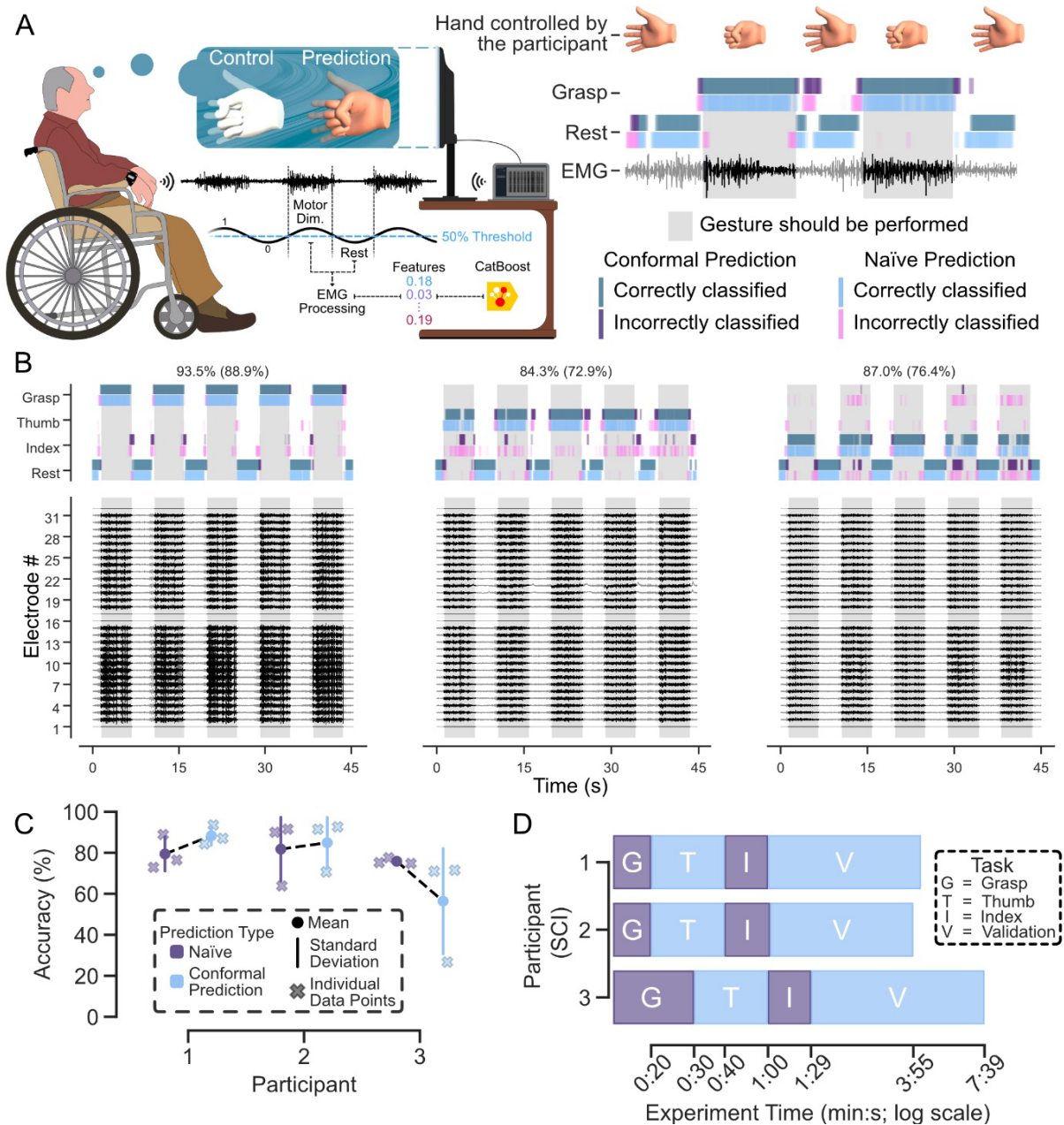

*Fig. 3. **Decoding of hand degrees of freedom for individuals suffering from a SCI.** (A) Three spinal cord injured participants (Table 1) were asked to follow three different movements shown on a screen. Using a 32-channel bracelet, their forearm EMG activity was streamed and processed by MyoGestic (see Materials and Methods). MyoGestic records the kinematics of the virtual hand as well as the synchronous EMG signals and identifies the motor commands during the attempted movements by the participants with neural lesion with a machine learning approach. We validated our system by asking the participants to follow the same trained movements again in real-time. To counteract EMG signal variability, we embedded in MyoGestic a conformal prediction algorithm (see Materials and Methods) to quantify and resolve prediction uncertainties. The participants followed a sinusoidal movement for each degree of freedom for 20 s (30 s for the third participant). To increase rest-state detection robustness despite spasticity, we defined a movement as performed when the guiding hand was past the 50% flexed state. Below 50%, we considered it as rest-state. Video S3 shows the third participant performing the experiment described before. (B) Exemplary real-time prediction and EMG signals of the first participant with SCI. The accuracy achieved with conformal prediction is displayed at the top of each movement, with the naïve approach shown in parentheses. (C) Classification accuracy for each participant using the naïve approach and conformal prediction (see Materials and Methods). The accuracies for the participants were 79.4 ± 8.4%, 81.8 ± 15.6%, and 75.8 ± 1.5% for the naïve approach, and 88.3 ± 4.7%, 84.9 ± 12.4%, and 56.4 ± 25.7% using conformal prediction. (D) The time taken for each participant to record and validate four degrees of freedom was 4 min 12 s, 3 min 55 s, and 7 min 39 s. We asked each participant to execute each movement for 20 seconds, and to perform 6 repetitions within a 45-second span for validation.*

**Motor intent decoding from individuals with a transradial amputation**

We also recorded data from two individuals with transradial amputation (Table 1; see Materials and Methods) using our neural interface system. Similar to the participants with SCI, our first goal was to determine how many degrees of freedom could be decoded from the bracelet signals in real-time and to assess their decoding speed. For the spinal cord injured patients, we focused on four degrees of freedom (resting state plus three movements). In contrast, we asked the transradial amputees to attempt to flex each digit individually.

For the amputees we recorded only the end state of each degree of freedom for 10 seconds (Fig. 4A). Both participants could differentiate between rest and four other movements. The first transradial amputee could not reliably produce different EMG signals to differentiate between the index and middle finger. The second amputee could not voluntarily control the pinky finger and could not imagine the thumb flexing independently of the index finger. For the thumb, we remapped intended thumb flexion as extension. In summary the first amputee could control the thumb, the index and middle fingers together, the ring finger, and the pinky. The second amputee could control every digit except the pinky. To the best of our knowledge, this is the first study showing a high number of degrees of freedom that can be controlled in real-time by individuals with a missing limb.

Fig. 4B shows the predictions synchronized with the EMG signals of the first amputee. Data from the second participant can be found in Fig. S2B. Video S4 shows the first amputee controlling five degrees of freedom in real-time. Fig. 4C shows the accuracy of the two amputees both with (52.2 ± 21.0% and 74.2 ± 15.7%) and without uncertainty awareness and solving (58.5 ± 9.7% and 73.5 ± 15.6%). The time taken to achieve five degrees of freedom was 6 min 4 s and 5 min 42 s, respectively (Fig. 4D).

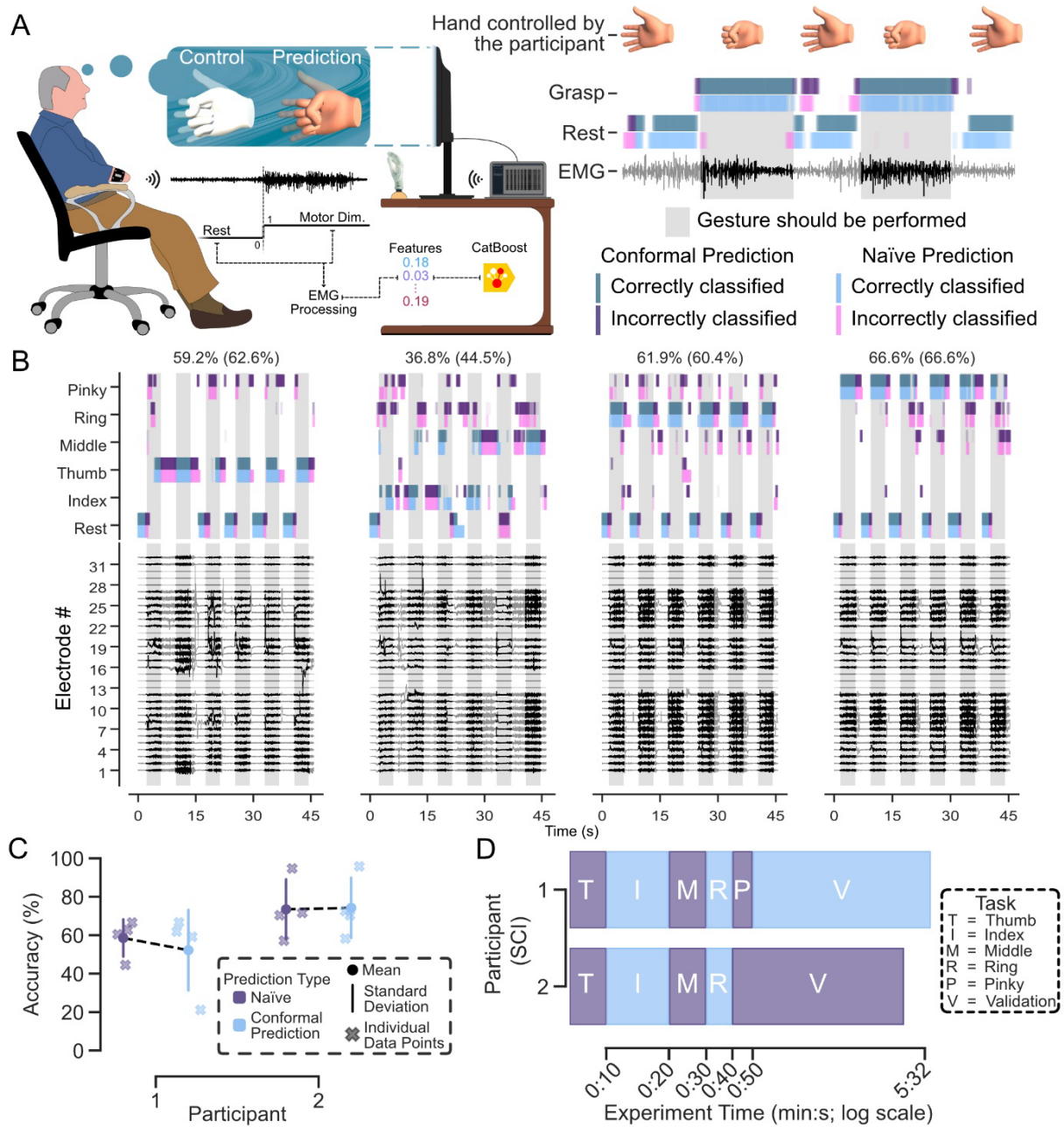

*Fig. 4. **Decoding of hand degrees of freedom for individuals suffering from a transradial amputation.** (A) Two transradial amputees (Table 1) were asked to flex all their fingers while being visually guided by a hand displayed on a screen. Using the proposed 32-channel bracelet, their forearm EMG activity was streamed and processed by our software, MyoGestic. We validated our system by having the participants repeat the same trained movements in real-time. To address EMG signal variability, we applied conformal prediction (see Materials and Methods) to quantify and resolve prediction uncertainties. The participants were instructed to hold the fully flexed state of each finger for 10 seconds. Video S4 shows the first participant performing the described experiment. (B) Exemplary real-time prediction and EMG signals of the first amputee. The accuracy achieved with conformal prediction is displayed at the top of each movement, with the naïve approach shown in parentheses. (C) Classification accuracy for both amputees using the naïve approach and conformal prediction (see Materials and Methods). The accuracies for the participants were 58.5 ± 9.7% and 73.5 ± 15.6% for the naïve approach, and 52.2 ± 21.0% and 74.2 ± 15.7% using conformal prediction. (D) Time taken for each participant to record and validate five degrees of freedom was 6 min 4 s and 5 min 42 s. We asked each participant to execute each movement for 10 seconds, and to perform 6 repetitions within a 45-second span for validation.*

**Adaptability to different experimental scenarios: bionic limb, active exoskeletons, and multidimensional cursor control**

To demonstrate the adaptability of our software framework, we tested it in five different and realistic scenarios that could be of interest to the myocontrol research community (Fig. 5 – scenario 1 to 4, Video S1 – scenario 5).

We integrated the Michelangelo Hand prosthesis (Ottobock, Duderstadt, Germany) as an output system into MyoGestic. One of our participants with transradial amputation controlled the bionic hand through MyoGestic in an effective and intuitive way (Fig. 5A). The prosthesis was placed on a socket in front of the participant (Fig. 4A; Fig. 5A). The participant was asked to follow the virtual guiding hand (Fig. 1; Fig. 2A), which showed a continuous opening and closing of a three-finger pinch, and to execute the displayed movements bilaterally for visual confirmation of execution.

A participant with a SCI affecting the lower limb (Table 1) was recruited to control a 2D cursor with five degrees of freedom (Fig. 5B). For this, we developed a larger EMG bracelet (33 cm) to record the activity of the calf muscles. We recorded the rest position, as well as the maximum possible inversion, eversion, dorsiflexion, and plantarflexion of the foot. Each movement was recorded starting from and returning to the rest position. For real-time cursor testing, we divided the movements into left-right and up-down directions to reduce cognitive strain on the participant. The movements always included a brief stop in the rest position before continuing in the opposite direction.

For the Michelangelo Hand, we relied on a preexisting communication protocol. However, we also implemented a connection from scratch for a custom-built hand orthosis (see Walter et al. (*22*) for details). We asked the first spinal cord injured participant to use the orthosis to attempt the box and block challenge (Fig. 5C). Additionally, we integrated a custom-built deep neural network (see Materials and Methods) into MyoGestic to predict hand position proportionally rather than just classifying the start and end states. Fig. 5D shows the first spinal cord injured participant attempting a proportional pinky movement. We chose the pinky finger because it was easy for the participant to control the EMG activity by attempting pinky movements.

Lastly, we recruited a transcarpal (forearm muscles are preserved) amputee (Table 1) to examine the possibility of decoding the motor intent of each finger (Video S1).

These tests suggest that MyoGestic is the first open-source software offering an intuitive and tailored machine learning framework, which helps individuals with various neural lesions to control their limb movements again.

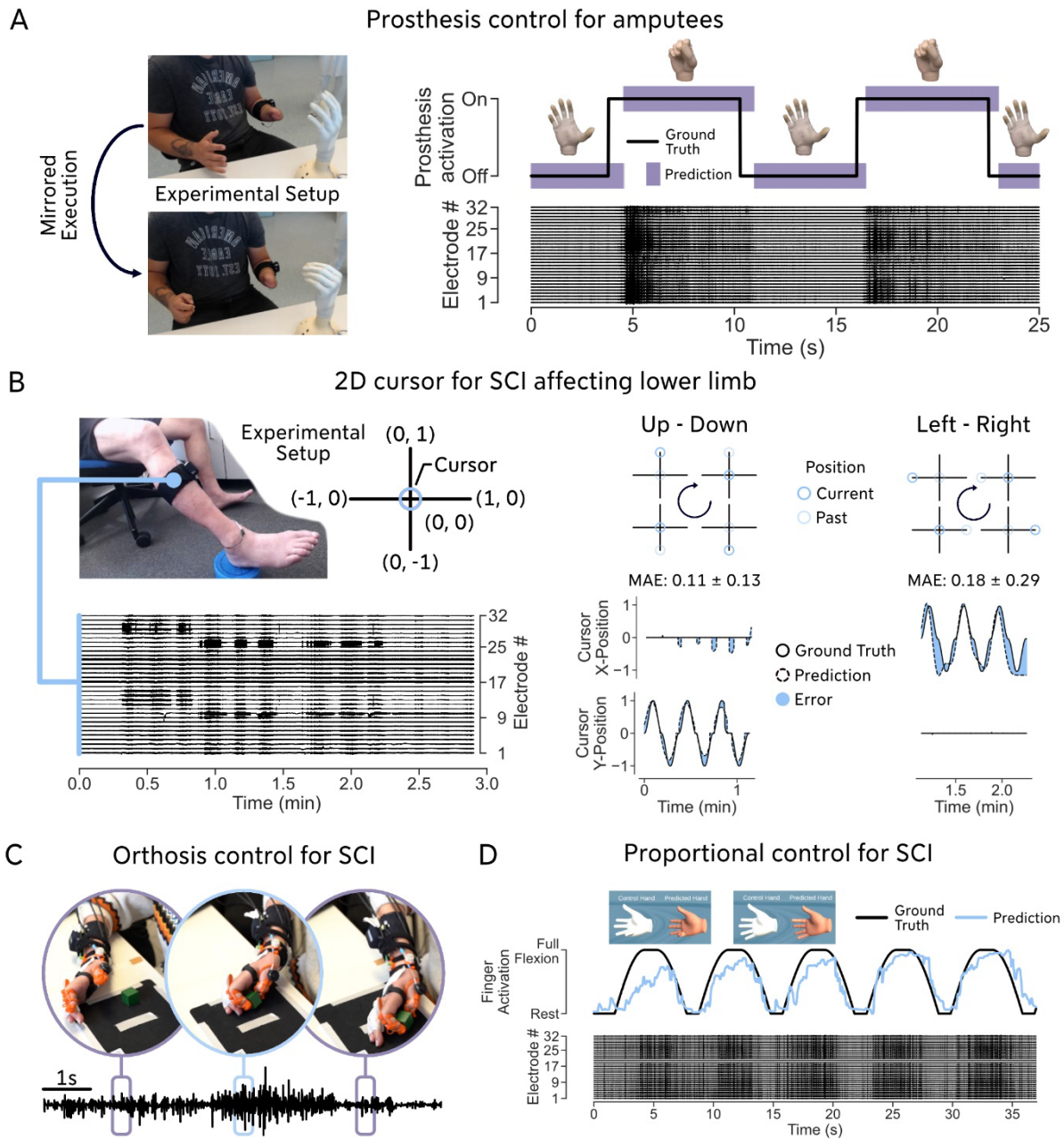

*Fig. 5. **Demonstration of MyoGestic's adaptability to different experimental scenarios.** Our software framework, MyoGestic, is designed to be agnostic to input, algorithm, and output. For user convenience, we maintain the EMG bracelet while varying the output system (A-C) and algorithms (D) to demonstrate its capabilities. **(A)** The second transradial amputated participant was asked to control the three-finger pinch of the Michelangelo Hand prosthesis (Ottobock, Duderstadt, Germany). The participant was instructed to perform the hand-closing motion with both hands for visual confirmation of the movement. To guide the participant, a visual interface was employed (Fig. 2B). **(B)** A fourth spinal cord injured participant (Table 1) was recruited to control a 2D cursor using the neural activity from their spared leg. The EMG bracelet was placed around the bellies of the calf muscles. The participant was instructed to perform inversion-eversion (left-right) and dorsiflexion-plantarflexion (up-down) movements, always pausing briefly in a natural rest position. **(C)** The first spinal cord injured participant was outfitted with a custom-built orthosis (see Walter et al. (22) for design details) and asked to perform the box and blocks test. **(D)** Lastly, we implemented a deep neural network (see Materials and Methods) to proportionally predict the control of a degree of freedom that the first spinal cord injured participant could best remember and envision.*

## DISCUSSION

Our study demonstrates the potential of our novel neural interface system called MyoGestic to decode motor intentions in two individuals with SCI, two with stroke, and three amputees.

### Motor intent decoding from individuals with SCI

We show that one participants with cervical SCI and two with spinal stroke (Table 1) could produce distinct EMG signals for grasping, flexing the thumb, and the index finger (Fig. 3), despite long-term paralysis. This finding reconfirms previous studies (*5*, *6*, *15*, *16*) indicating the persistence of some motor units embedding clear EMG patterns even after SCI and prompts a reevaluation of current SCI assessment standards (*9*), suggesting they may need to be extended to incorporate EMG signal evaluation.

In our previous work, we demonstrated that a deep-learning system could identify which movements are proportionally decodable in participants with motor-complete SCI (*16*). We proposed that such techniques could enhance injury assessment by distinguishing between EMG-complete and EMG-incomplete statuses. This study builds on our previous research by showing that voluntary signal assessment can be accomplished in under 10 minutes (Fig. 3D) using comparatively simple algorithms (see Materials and Methods). Our open-source framework, MyoGestic (Fig. 2; Fig. S1A) could thus aid medical professionals in determining the presence of voluntary EMG signals in individuals with SCI. This capability not only has the potential to positively influence the psychological state of affected individuals but also supports researchers and medical professionals in designing and selecting the most suitable assistive devices.

### Motor intent decoding from individuals with a transradial amputation

We tested our neural interfacing system on two transradial amputees and found that the EMG signal had some artifacts (Fig. 4B), primarily because our bracelets, although somewhat form-compliant, were not perfectly tight around the participants' stumps. This issue could be resolved by tailoring a participant-specific bracelet with distinct sizes and shapes.

Despite this challenge, we demonstrated that we could decode and validate five degrees of freedom in under 10 minutes (Fig. 4) using MyoGestic. Additionally, the typical control paradigm for prosthetics involves using muscle cocontractions to change the selected movement patterns, which is not intuitive for the user. MyoGestic allows for intuitive control by enabling the user to e.g., "actually grasp and not just initiate the grasping action by cocontracting" (reported by one of our participants).

### Generalizable control

MyoGestic was designed as a versatile interface to connect the motor intent of participants with various output modalities. Our study demonstrated that six participants, three suffering from SCI and three from amputations (Table 1), successfully controlled a virtual hand (Fig. 3; Fig. 4).

Additionally, we showed in 5 different cases realistic use cases of MyoGestic:

- connecting to a commercially available system using a preexisting communication protocol (Fig. 5A),
- interfacing the lower limb as opposed to the upper one and outputting the decoded degrees of freedom as a 2D cursor as opposed to a virtual hand (Fig. 5B),
- interfacing with a custom-build system and thus requiring a new implementation of a communication protocol from scratch (Fig. 5C),

- integrating a new algorithm that can do regression instead of classification (Fig. 5D),
- and controlling all hand digits using the preserved forearm muscles after amputation (Video S1).

These results indicate the potential of MyoGestic as a universal interface between users and diverse output devices. Future research should explore the extension of MyoGestic's capabilities to help other relevant problems, such as integrating with neuroorthoses for enhancing or restoring mobility.

**EMG Hardware - Comparison with commercial standards**

Lastly, it is important to highlight the difference between our system and commercial assistive devices.

Commercial assistive devices are tailored to the shape and size of the patient anatomy, while our hardware was not. To accommodate a broad participant population, we produced five bracelet sizes (18, 19, 21, 23, and 33 cm). This is particularly important for amputees, as our bracelets cannot perfectly conform to the preserved anatomy, resulting in noisy channels due to suboptimal skin-electrode contact (Fig. 4B). For optimal results, each participant should ideally be measured, and a custom bracelet should be designed for them. A custom fit would likely improve the accuracy and reliability of the EMG signal recordings.

Most assistive devices use a low number of electrodes (fewer than 10), which, as our previous research (*23*) shows, negatively impacts hand kinematics predictions. In contrast, our system uses 32 electrodes, allowing it to decode richer patterns and achieve an unprecedented level of degrees of freedom (Fig. 3, Fig. 4).

The last significant difference is the lack of dedicated learning time with our neural interface for the participants. Unlike most assistive devices that necessitate extensive physiotherapy and rehabilitation sessions to master the system, our participants did not have such preparatory sessions, in which potentially more data can be recorded per movement compared to our maximum of 30 s. While the lack of training could impact the participants' proficiency in generating clear and consistent EMG signals, potentially affecting the system's performance, it also demonstrates that executing intuitive movements is preserved even after years of limb function loss.

**MATERIALS AND METHODS**

**Study design**

We recruited two participants with a SCI, two with stroke, and three with an amputation (Table 1). Participants gave their informed written consent to participate in the study. The study was carried out in accordance with the Declaration of Helsinki. All procedures and experiments were approved by the ethics committees of the Friedrich-Alexander-Universität Erlangen-Nürnberg (applications 22-138-Bm and 21-150-B) and of the Eberhard-Karls-Universität Tübingen (applications 470/2019B02 and 181/2020B01). Before the experiments, the participants received a detailed explanation of the study again and were asked to reaffirm their consent to participate.

*Table 1. **Participant characteristics.** Age is given in 5-year ranges to protect the participants privacy. IL = Injury Level. ZPP = Zone of Partial Preservation (9). AIS = ASIA Impairment Scale (9).*

| Participant ID | Experiment taken part in (all in real-time) | Injury type | Time since injury (years) | Age (5-year range) | Sex |
|---|---|---|---|---|---|
| SCI 1 | Control of a virtual hand (Fig. 3) Orthosis control (Fig. 5C) Proportional control (Fig. 5D) | spinal cord ischemia C5 / S1 / D (IL / ZPP / AIS) | 12 | 55 - 60 | F |
| SCI 2 | Control of a virtual hand (Fig. 3) | spinal cord ischemia C5 / S5 / D (IL / ZPP / AIS) | 2 | 50 - 54 | M |
| SCI 3 | Control of a virtual hand (Fig. 3, Video S3) | traumatic spinal cord injury C6 / C7 / B (IL / ZPP / AIS) | 21 | 40 - 44 | M |
| SCI 4 | 2D foot cursor (Fig. 5B) | traumatic spinal cord injury L3 / S1 / D (IL / ZPP / AIS) | 3 | 60 - 64 | M |
| Amputee 1 | Control of a virtual hand (Fig. 4, Video S4) | transradial amputation | 1 | 45-50 | M |
| Amputee 2 | Control of a virtual hand (Fig. 4) Prosthesis control (Fig. 5A) | transradial amputation | 0.42 | 20-24 | M |
| Amputee 3 | Hand digit control (Video S1) Demonstration of workflow (Video S2) | transcarpal amputation | 0.92 | 35-40 | F |

Each experiment was conducted in a single session per participant. The initial 10 minutes were allocated for hardware setup and introduction to the virtual hand interface (Fig. 1, Fig. 2) or the 2D cursor for the leg experiments (Fig. 5B). For participants suffering from a SCI, the bracelet was placed 3 to 5 cm above the distal ulnar head for the participants convenience (Fig. 2C). This position was chosen because it replicated the position of a wristwatch, something most participants use daily, thus minimizing the need for adjustment. We placed the bracelet on the amputees as close as possible to the site of amputation on the remaining muscle bellies under the guidance of the presiding medical professional. For the fourth participant suffering from SCI, we placed the bracelet on the bellies of the calf muscles (Fig. 5B). Five bracelet lengths (18, 19, 21, 23, and 33 cm) were available, with the one closest to the participant's forearm (or leg) circumference selected to ensure optimal EMG signal quality and minimal electrical noise interference. An adhesive ground reference electrode was attached to the elbow or the medial malleolus for the leg.

The next 15 minutes were dedicated to identifying the hand (foot for SCI 4) movements the participants could reliably recall after years of living with motor function loss. The following movements could be displayed on a screen in front of them as a virtual hand to provide visual aid (Fig. 1, middle):

- thumb flexion,
- index flexion,
- middle flexion,
- ring flexion,
- pinky flexion,

- grasp,
- thumb and index pinch, and
- thumb, index, and middle finger pinch.

For the foot we used a red circle that could move up, down, left, and right as a 2D cursor (Fig. 5B).

We used MyoGestic's real-time plotting capabilities to visually inspect the participants' recalled movements, ensuring that they were distinguishable from each other. Our primary goal was to maintain intuitiveness by having participants attempt to execute and be able to display the same executed movement on the screen. However, we remained adaptable in case the prepared movements were not recallable or distinguishable. MyoGestic allowed us to map the executed movements to alternate ones for display without any code changes. For example, if a participant recalled how to extend their fingers instead of how to flex them, we could use this as a proxy for grasping. In this case, the grasping action would still be accurately displayed in the flexing state (Fig. 1, middle).

After identifying different movements (three for SCI and four for amputees) that are both visually distinguishable and reliably recallable by the participant, we displayed each of said movements in a sinusoidal pattern (relaxed state to fully executed) on the display at a speed of 1 execution per 5 s with 1.5 s hold time for both the rest phase and the executed phase. The EMG together with the movement kinematics were recorded for 30 s per movement. To counteract spasticity, which unattended results in poor rest-state behavior, we separated every movement task into an active state and a rest state. This ensured that the rest state could be reached even if the spastic MU remained active long after the movement was executed, and the participant's hand returned to a resting position. We chose 50% movement activation for the class boundary. In total the rest state had 45 s of data, while each other movement had 15 s each.

**Bracelet**

The EMG bracelet consisted of 32 gold-coated copper electrodes spread over 2 columns with 16 rows. Column-wise, the interelectrode distance (IED) is fixed at 2 cm, while row-wise it varied between 1 and 1.5 cm depending on the bracelet length selected. The bracelet together with the amplifier weighed $76 \pm 2$ g (measured using a kitchen scale, OK. OKS 3220). Exact measurements can be found in Fig. S3.

The raw signals from the electrodes were sampled at 2000 Hz, amplified with a gain of 4, digitized to 16 bits, and streamed at approximately 111 Hz by a commercially available wireless amplifier (MUOVI, OT Bioelettronica S.r.l., Turin, Italy). The amplifier thus could stream 111 non-overlapping EMG matrices (32 channels x 18 samples) per second of data via a Transmission Control Protocol (TCP) connection.

**Visual interface**

The visual interface built using Unity 2021.3.20f1 (Unity Technologies, San Francisco, USA) displayed two hands: a *control hand*, which controlled the user's movement pace, and a *predicted hand*, which provided real-time feedback on the movements attempted by the participants (Fig. 1). The hand model (HTC VIVE Unity SDK, HTC Corporation, New Taipei, Taiwan) was animated using a bone rig that allows each hand joint to be represented by a quaternion.

To aid the participants in imagining and following a movement, the control hand could display nine distinct hand movements: a rest state, individual digit flexions, a grasp, and a two-finger (thumb and index) or three-finger (thumb, index, and middle) pinch. The frequency and holding time for both extrema (resting and full flexion) could be adjusted to best accommodate each participant. For our experiments we set the holding time to be 1.5 s resulting in 1 movement cycle (resting to full flexion and back) per approximately 7.5 s. The control hand was represented in any state by a 9D vector with values ranging between 0 and 1 (Fig. 2B). The rest state was represented by a vector of nine 0 s. The first two values control thumb flexion and abduction, respectively. The next four values control the flexion of the index, middle, ring, and pinky fingers. The last three values represent wrist flexion, adduction, and pronation. During recording, both for training and validating the model, the control hand state was streamed and saved at 60 Hz using a User Datagram Protocol (UDP) connection.

The prediction hand could also be represented by the same 9D vector structure to provide feedback to the participant. For the classification approach in this work, we use linear interpolation between the quaternions described by the previous and new vector states to create a smooth transition rather than abruptly jumping from rest to a closed grasp, for instance. The state of the prediction hand was set via UDP at approximately 32 Hz.

**Software interface**

To facilitate a smooth development experience for future work, we built our software framework in Python. This framework is designed to be flexible and accommodate a wide range of experiments by providing only the necessary components such as a graphical user interface (GUI) template that can be edited and extended, real-time plotting capabilities, and pre-built connection templates for various I/O solutions. Video S2 shows in real-time the workflow necessary for setting up and recording three degrees of freedom from the third amputee.

For this work, we subdivided the software interface into three distinct panels: input, output, and algorithm/processing. Each panel could only be displayed one at a time. This design choice was made to ensure that users could focus on fewer steps at a time and minimize human errors during hospital visits.

The first panel was used to connect to the Wi-Fi amplifier and to set the recording configurations (see Bracelet section for details). Once connected, the signals were displayed in real-time on the right side of the interface, allowing the user to adjust the bracelet if necessary. Connection to the visual interface via UDP was established in the second panel. The last panel contained all necessary functionality to record training data, train, and validate the myocontrol algorithm. The GUI for recording training data allowed users to select the recording duration and choose the movements they wished to display when executing their chosen movements. The displayed movement could either match the executed one or differ, depending on how well the participant remembered a particular movement. We chose 30 s per movement and selected the movements after discussing with each participant which movements they could remember. Generally, we found that grasping and flexing the index finger were not forgotten by the participants, while flexing the pinky finger individually was often imagined as flexing the ring and pinky fingers together.

The wireless amplifier streamed the EMG signals at 111 Hz and recorded them at 2000 Hz. Therefore, for 1 s of recording, the amplifier streamed 111 matrices of 32 EMG channels with 18 samples per channel (9 ms). However, as 9 ms was not enough temporal information, we created a real-time queue that buffered 20 EMG windows, resulting in a total signal length of approximately 180 ms (360 samples). The 32 EMG channels have been bandpass filtered between 10 and 500 Hz at base by the amplifier.

To suppress artifacts and other undesired noise, the 32 channels were restructured into their physical configuration of 16 rows by 2 columns (Fig. 2E). Given that the bracelet forms a loop, the 16 x 2 EMG matrices were circularly padded row-wise: the last row was added on top of the first row, and the first row was appended below the last row to simulate the physical arrangement. Each column was zero-padded to facilitate the application of a 3 x 3 filter across the entire grid. A 3 x 3 blur filter was employed to suppress high-frequency noise:

$$\frac{1}{4}\begin{pmatrix} 0 & 1 & 0 \\ 1 & 0.5 & 1 \\ 0 & 1 & 0 \end{pmatrix} \quad (1)$$

This filter, at any given time, could only consider the current electrode and its three immediate neighbors. One RMS value was then computed for each channel and saved together with the mean of the control hand state (Fig. 2B). This resulted in 32 features that corresponded to only one 9D hand state vector.

From experience, we knew that reaching the rest state after attempting a movement is difficult for individuals with SCI due to spasticity (*24*), which involves involuntary overactivity of muscles. To address this, we defined the rest state as occurring when the control hand was below the 50% mark of the fully flexed state for any movement. Anything at or above 50% was then classified as movement. The first 80% of the data was allocated for training, with the remaining 20% reserved for testing. Mean and standard deviation were computed from the training data and used to normalize both the testing set and all data used for the real-time prediction. Using this training dataset, we employed a CatBoost model (*19*) trained on an NVIDIA RTX 4090 Mobile graphical processing unit. Training lasted for 1000 epochs, resulting in approximately $5.21 \pm 0.12$ s (computed with 120 s of EMG data for n = 1000 runs) of training time.

**Conformal prediction**

Surface EMG signals are inherently variable (*8*) because of physiological factors such as motor units rotating between actively participating in movement generation and resting to avoid fatigue (*25*), the amount of subcutaneous fat layer thickness a participant has (*26*) or the amount of cross-talk present. In individuals with SCI further variability is introduced in the form of spasticity (*24*).

Conformal prediction is a statistical framework that provides a proxy measurement for uncertainty (*27*). By generating prediction intervals or sets that contain the true outcome with a specified probability, known as the confidence level, it offers more insight into the uncertainty of the model. The size of the prediction set can be used to gauge the model's uncertainty. In this work, a set containing only one prediction was considered certain. We used the conformal prediction approach of regularized adaptive prediction sets (RAPS). This method avoids the prediction of empty sets while maintaining a regulated set size due to an applied penalty weighting.

To test whether uncertainty awareness would minimize error and improve the predictions for the participants, we created a real-time capable temporal uncertainty solver that we ran in simulated real-time after the experiments. Sets containing more than one possible prediction were filtered by considering the last 75 sets. To derive a single prediction, we counted the occurrence of each prediction within the involved sets and determined the most frequent prediction as the output.

**Deep regression neural network**

We adapted our previously published deep learning model (*20*, *21*) to enable proportional control of hand degrees of freedom instead of classification. The original model was a 3D convolutional neural network (CNN) that utilized five 64-channel grids. Since the bracelet is a 32-channel grid, we modified the network from a 3D CNN to a 2D CNN, keeping the rest of the architecture unchanged.

**List of Supplementary Materials**

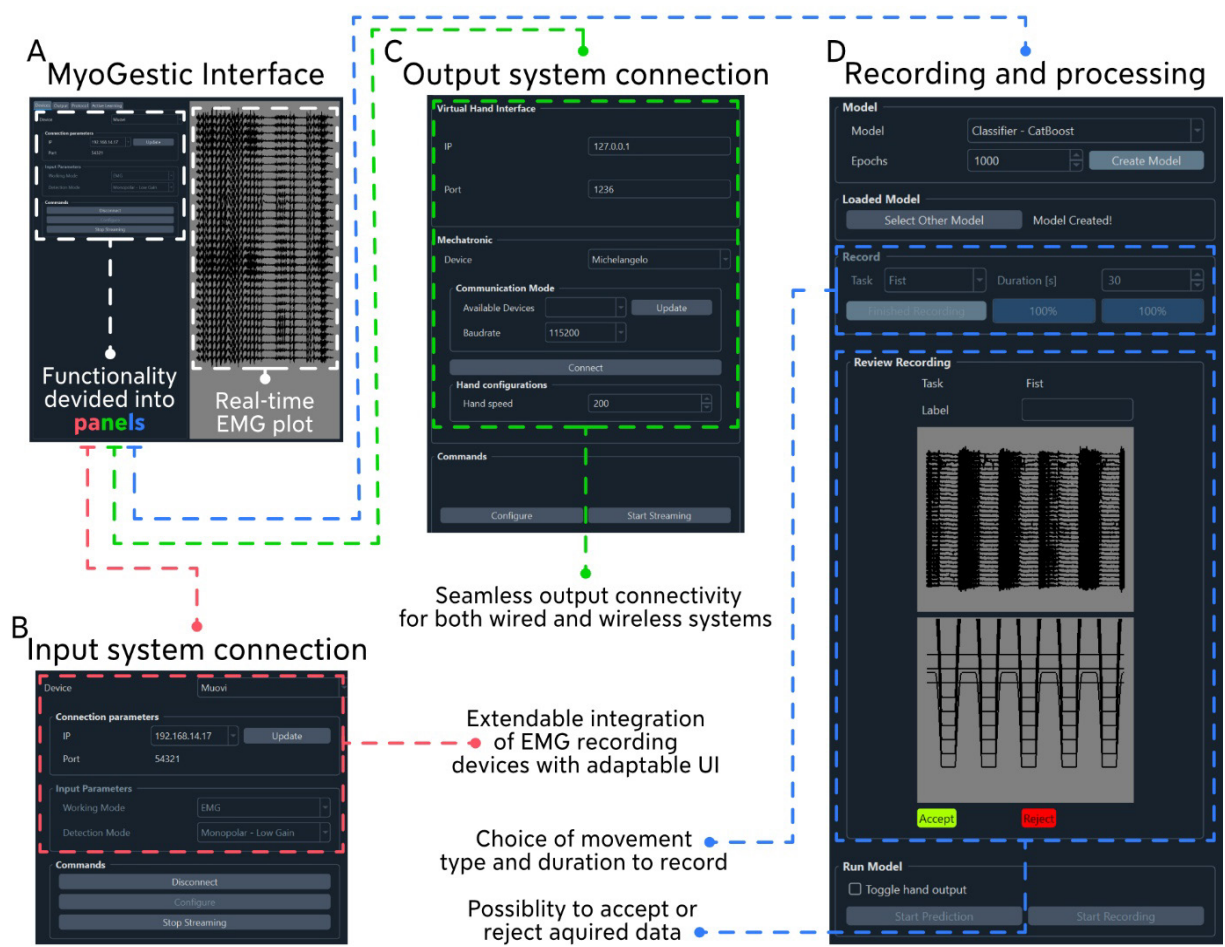

*Fig. S1. **MyoGestic interface overview and functionality highlights.** (**A**) The MyoGestic interface streamlines experimentation by guiding both experimenter and participant through each step, providing transparency on current progress and objectives. The functionality (left side) is divided into three main panels: input system (B), output system (C), and recording and processing (D). Each panel can only be displayed one at a time to minimize human error and in turn participant frustration. On the right side, all 32 EMG channels are displayed in real-time, allowing for immediate identification of noise sources to ensure optimal signal quality. (**B**) The first panel connects to the input device, such as our wireless EMG bracelet. The framework is designed to be adaptable and can accommodate any biosignal acquisition device, providing an intuitive solution for integrating the necessary configuration into the user interface (UI). (**C**) The second panel connects to the output device. Just as the first panel any software of hardware output system can be accommodated. (**D**) The final panel manages the experiment workflow, encompassing recording, processing, and validation of myocontrol algorithms. We designed it for ease-of-use, allowing users to record new movements and re-record if necessary (e.g., due to noise or participant errors). This design allows fast accommodation of any participant feedback.*

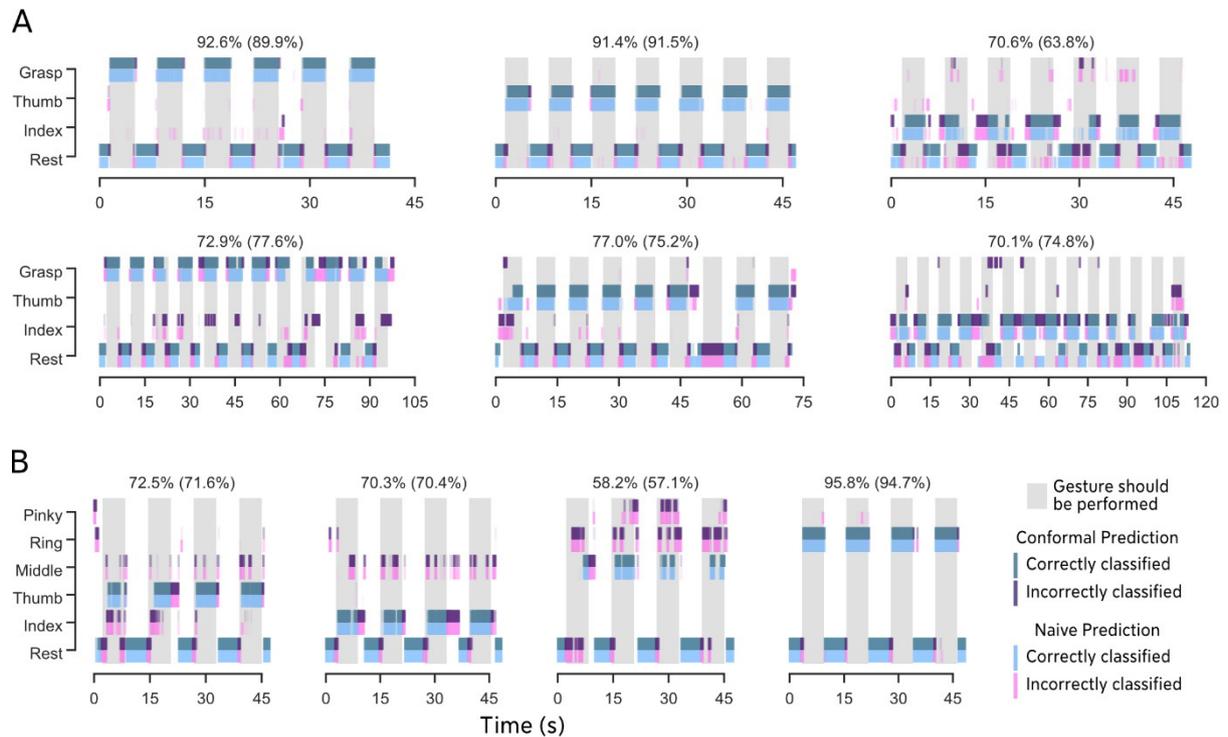

*Fig. S2. **Results from the experiments in Fig. 3 & Fig. 4 for the other participants not shown.** (A) The results for the second and third spinal cord injured participant (Table 1) from the experiment described in Fig. 3. (B) The results for the second transradial amputee (Table 1) from the experiment described in Fig. 4.*

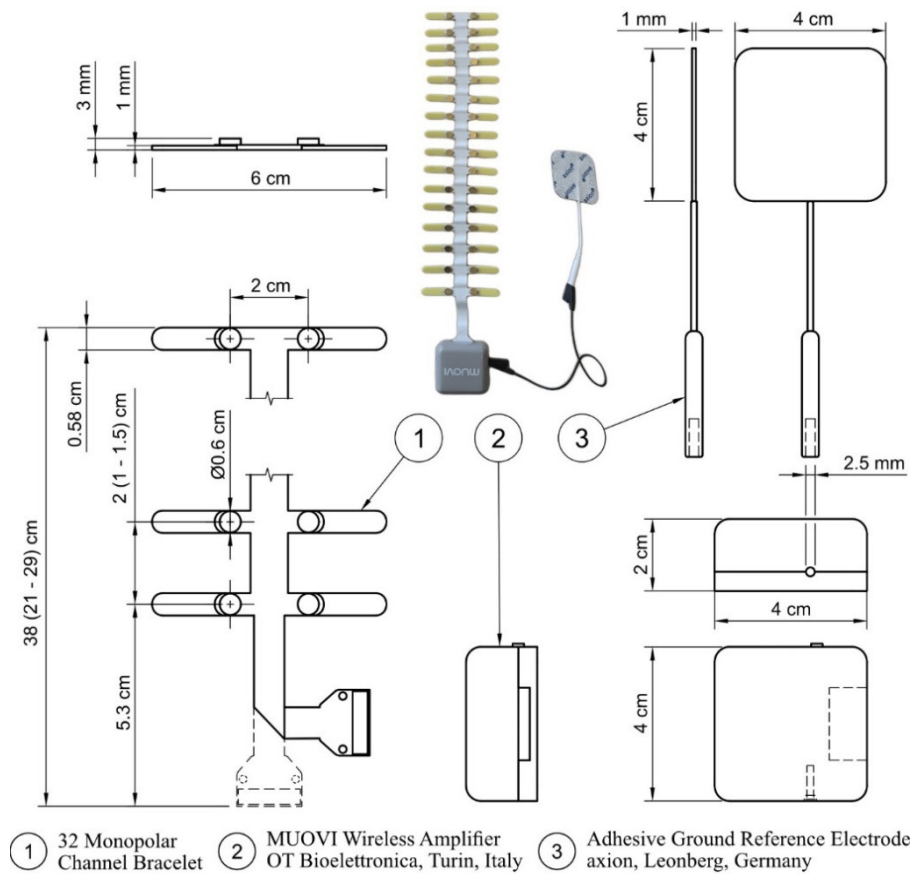

*Fig. S3. **Technical schematic of the bracelet components.** The 32 electrodes (①, without fabric sleeve) are arranged in a 16 x 2 configuration affixed on a thin PCB and connected row-wise by a flexible cable. The commercially available Wi-Fi EMG amplifier (②, MUOVI, OT Bioelettronica S.r.l., Turin, Italy) is attached by a custom connector to the electrodes. The amplifier*

*can then be stored in a fabric pocket on the bracelet for ease of wear. The commercially available ground reference (③, Adhesive electrode 4x4 cm, axion GmbH, Leonberg, Germany) is placed on a bone (e.g., elbow) for ideal signal quality and is attached to the amplifier by a 2.5 mm jumper wire.*

*Video S1 Real-time full digit control using MyoGestic executed by the transcarpal amputee.*
*Link: https://youtu.be/NPemwlSg-mE*

*Video S2 Annotated depiction of the MyoGestic workflow, illustrating the process from software initialization to the control of 3 degrees of freedom in a transcarpal amputee.*
*Link: https://youtu.be/vdP5Ci9cvR4*

*Video S3 Real-time control of 4 degrees of freedom executed by the third spinal cord injured participant.*
*Link: https://youtu.be/3BvVAu8Nq8c*

*Video S4 Real-time control of 5 degrees of freedom executed by the first transradial amputee.*
*Link: https://youtu.be/zxICSVn-3P8*

**References and Notes**

1. W. Ding, S. Hu, P. Wang, H. Kang, R. Peng, Y. Dong, F. Li, Spinal cord injury: the global incidence, prevalence, and disability from the global burden of disease study 2019. *Spine* **47**, 1532 (2022).

2. M. Giroud, A. Jacquin, Y. Béjot, The worldwide landscape of stroke in the 21st century. *Lancet* **383**, 195–197 (2014).

3. F. Romi, H. Naess, Spinal cord infarction in clinical neurology: a review of characteristics and long-term prognosis in comparison to cerebral infarction. *Eur. Neurol.* **76**, 95–98 (2016).

4. C. L. McDonald, S. Westcott-McCoy, M. R. Weaver, J. Haagsma, D. Kartin, Global prevalence of traumatic non-fatal limb amputation. *Prosthet. Orthot. Int.* **45**, 105 (2021).

5. J. E. Ting, A. Del Vecchio, D. Sarma, N. Verma, S. C. Colachis, N. V. Annetta, J. L. Collinger, D. Farina, D. J. Weber, Sensing and decoding the neural drive to paralyzed muscles during attempted movements of a person with tetraplegia using a sleeve array. *J. Neurophysiol.* **126**, 2104–2118 (2021).

6. D. Souza de Oliveira, M. Ponfick, D. I. Braun, M. Osswald, M. Sierotowicz, S. Chatterjee, D. Weber, B. Eskofier, C. Castellini, D. Farina, T. M. Kinfe, A. Del Vecchio, A direct spinal cord–computer interface enables the control of the paralysed hand in spinal cord injury. *Brain*, awae088 (2024).

7. D. Farina, I. Vujaklija, R. Brånemark, A. M. J. Bull, H. Dietl, B. Graimann, L. J. Hargrove, K.-P. Hoffmann, H. (Helen) Huang, T. Ingvarsson, H. B. Janusson, K. Kristjánsson, T. Kuiken, S. Micera, T. Stieglitz, A. Sturma, D. Tyler, R. F. ff Weir, O. C. Aszmann, Toward higher-performance bionic limbs for wider clinical use. *Nat. Biomed. Eng.* **7**, 473–485 (2023).

8. R. Merletti, D. Farina, Eds., *Surface Electromyography : Physiology, Engineering, and Applications* (Wiley, ed. 1, 2016).

9. R. Rupp, F. Biering-Sørensen, S. P. Burns, D. E. Graves, J. Guest, L. Jones, M. S. Read, G. M. Rodriguez, C. Schuld, K. E. Tansey-Md, K. Walden, S. Kirshblum, International standards for neurological classification of spinal cord injury: revised 2019. *Top. Spinal Cord Inj. Rehabil.* **27**, 1–22 (2021).

10. I. Campanini, C. Disselhorst-Klug, W. Z. Rymer, R. Merletti, Surface EMG in clinical assessment and neurorehabilitation: barriers limiting its use. *Front. Neurol.* **11** (2020).

11. CTRL-labs at Reality Labs, D. Sussillo, P. Kaifosh, T. Reardon, A generic noninvasive neuromotor interface for human-computer interaction. bioRxiv [Preprint] (2024). https://doi.org/10.1101/2024.02.23.581779.

12. A. Moin, A. Zhou, A. Rahimi, A. Menon, S. Benatti, G. Alexandrov, S. Tamakloe, J. Ting, N. Yamamoto, Y. Khan, F. Burghardt, L. Benini, A. C. Arias, J. M. Rabaey, A wearable biosensing system with in-sensor adaptive machine learning for hand gesture recognition. *Nat. Electron.* **4**, 54–63 (2020).

13. F. Chamberland, É. Buteau, S. Tam, E. Campbell, A. Mortazavi, E. Scheme, P. Fortier, M. Boukadoum, A. Campeau-Lecours, B. Gosselin, Novel wearable HD-EMG sensor with shift-robust gesture recognition using deep learning. *IEEE Trans. Biomed. Circuits Syst.* **17**, 968–984 (2023).

14. R. J. Varghese, M. Pizzi, A. Kundu, A. Grison, E. Burdet, D. Farina, Design, fabrication and evaluation of a stretchable high-density electromyography array. *Sensors* **24**, 1810 (2024).

15. D. I. Braun, D. S. de Oliveira, P. Bayer, M. Ponfick, T. M. Kinfe, A. D. Vecchio, NeurOne: high-performance motor unit-computer interface for the paralyzed. medRxiv [Preprint] (2023). https://doi.org/10.1101/2023.09.25.23295902.

16. R. C. Sîmpetru, D. S. de Oliveira, M. Ponfick, A. Del Vecchio, Identification of spared and proportionally controllable hand motor dimensions in motor complete spinal cord injuries using latent manifold analysis. medRxiv [Preprint] (2024). https://doi.org/10.1101/2024.05.28.24307964.

17. M. Nowak, I. Vujaklija, A. Sturma, C. Castellini, D. Farina, Simultaneous and proportional real-time myocontrol of up to three degrees of freedom of the wrist and hand. *IEEE Trans. Biomed. Eng.* **70**, 459–469 (2023).

18. C. Piazza, M. Rossi, M. G. Catalano, A. Bicchi, L. J. Hargrove, Evaluation of a simultaneous myoelectric control strategy for a multi-DoF transradial prosthesis. *IEEE Trans. Neural Syst. Rehabil. Eng.* **28**, 2286–2295 (2020).

19. L. Prokhorenkova, G. Gusev, A. Vorobev, A. V. Dorogush, A. Gulin, "CatBoost: unbiased boosting with categorical features" in *Proceedings of the 32nd International Conference on Neural Information Processing Systems* (Curran Associates Inc., Red Hook, NY, USA, 2018)*NIPS'18*, pp. 6639–6649.

20. R. C. Sîmpetru, M. März, A. Del Vecchio, Proportional and simultaneous real-time control of the full human hand from high-density electromyography. *IEEE Trans. Neural Syst. Rehabil. Eng.* **31**, 3118–3131 (2023).

21. R. C. Sîmpetru, A. Arkudas, D. I. Braun, M. Osswald, D. Souza de Oliveira, B. Eskofier, T. M. Kinfe, A. Del Vecchio, Learning a hand model from dynamic movements using high-density EMG and convolutional neural networks. *IEEE Trans. Biomed. Eng.*, 1–12 (2024).

22. J. Walter, P. Roßmanith, D. Souza de Oliveira, S. Reitelshöfer, A. Del Vecchio, J. Franke, "Proportional control of a soft cable-driven exoskeleton via a myoelectrical interface enables force-controlled finger motions" in *2022 9th IEEE RAS/EMBS International Conference for Biomedical Robotics and Biomechatronics (BioRob)* (2022; https://doi.org/10.1109/BioRob52689.2022.9925334), pp. 1–6.

23. R. C. Sîmpetru, V. Cnejevici, D. Farina, A. Del Vecchio, Influence of spatio-temporal filtering on hand kinematics estimation from high-density EMG signals. *J. Neural Eng.* **21** (2024).

24. F. Biering-Sørensen, J. B. Nielsen, K. Klinge, Spasticity-assessment: a review. *Spinal Cord* **44**, 708–722 (2006).

25. P. Bawa, C. Murnaghan, Motor unit rotation in a variety of human muscles. *J. Neurophysiol.* **102**, 2265–2272 (2009).

26. D. Souza de Oliveira, A. Casolo, T. G. Balshaw, S. Maeo, M. B. Lanza, N. R. W. Martin, N. Maffulli, T. M. Kinfe, B. M. Eskofier, J. P. Folland, D. Farina, A. D. Vecchio, Neural decoding from surface high-density EMG signals: influence of anatomy and synchronization on the number of identified motor units. *J. Neural Eng.* **19**, 46029 (2022).

27. A. N. Angelopoulos, S. Bates, M. Jordan, J. Malik, "Uncertainty sets for image classifiers using conformal prediction" in *Proceedings of the International Conference on Learning Representations (ICLR) 2021* (2021; https://openreview.net/forum?id=eNdiU_DbM9).

**Acknowledgments:**

**Funding:**

    European Research Council (ERC) grant 101118089 (ADV)

    German Ministry for Education and Research (BMBF) grant 01DN2300 (ADV)

    German Ministry for Education and Research (BMBF) grant 16SV9246 (ADV)

    German Research Foundation (DFG) grant 523352235 (ADV)

    Bavarian Ministry of Economic Affairs, Regional Development and Energy (StMWi) grant MV-2303-0006 (ADV)

    Bavarian Ministry of Economic Affairs and Media, Energy and Technology grant LSM-2303-0003 (ADV)

**Author contributions:**
    Conceptualization: RCS, DIB, ADV
    Data curation: RCS, DIB, MM, AUS, VC, DH,
    Formal Analysis: RCS, AUS
    Funding acquisition: DIB, ADV
    Investigation: RCS, DIB, AUS, DH
    Methodology: RCS, DIB, MM, AUS
    Project administration: RCS, DIB, ADV
    Resources: RCS, DIB, DSO, NW, JW, CP, MP, ADV
    Software: RCS, DIB, MM, AUS, VC
    Supervision: JF, CP, MP, ADV
    Validation: RCS, DIB, AUS

      Visualization: RCS
      Writing – original draft: RCS
      Writing – review & editing: RCS, DIB, AUS, VC, DSO, ADV

**Competing interests:** RCS, DIB, NW, JW, JF, and ADV have a patent submitted based on the orthosis described in this work.

**Data and materials availability:** See code at https://github.com/NsquaredLab/MyoGestic.